\documentclass{PoS}

\title{Hamiltonian approach\\ to Coulomb gauge Yang-Mills Theory}

\ShortTitle{Yang-Mills Theory in Coulomb gauge}

\author{\speaker{H.\ Reinhardt}\thanks{This work
was supported by the Deutsche Forschungsgemeinschaft (DFG) under contract no.\
Re856/6-1,2.}\\
        Institut f\"ur Theoretische Physik, 
Auf der Morgenstelle 14\\
D-72076 T\"ubingen, 
Germany\\
        E-mail: \email{h.reinhardt@uni-tuebingen.de}}

\author{W.\ Schleifenbaum\\
        Institut f\"ur Theoretische Physik, 
Auf der Morgenstelle 14\\
D-72076 T\"ubingen, 
Germany\\}

\author{D.\ Epple \\
        Institut f\"ur Theoretische Physik, 
Auf der Morgenstelle 14\\
D-72076 T\"ubingen, 
Germany\\}

\author{C.\ Feuchter\\
        Institut f\"ur Theoretische Physik, 
Auf der Morgenstelle 14\\
D-72076 T\"ubingen, 
Germany\\}

\abstract{The vacuum wave functional of Coulomb gauge Yang-Mills
  theory is determined within the variational principle and used to
  calculate various Green functions and observables. The results
  show that heavy quarks
  are confined by a linearly rising potential and gluons cannot
  propagate over large distances. The 't Hooft loop shows a perimeter
  law and thus also indicates confinement.}

\FullConference{The XXV International Symposium on Lattice Field Theory\\
		 July 30-4 August 2007\\
		 Regensburg, Germany}

\renewcommand{\phi}{\varphi}
\renewcommand{\epsilon}{\varepsilon}

\newcommand{\bee}{\begin{eqnarray}}
\newcommand{\eeq}{\end{eqnarray}}

\newcommand{\cN}{{\cal N}}

\newcommand{\cD}{{\cal D}}

\newcommand{\cA}{{\cal A}}

\newcommand{\cJ}{{\cal J}}

\newcommand{\be}{\begin{eqnarray}}
\newcommand{\ee}{\end{eqnarray}}
\newcommand{\hk}{\hspace{0.1cm}}

\newcommand{\rk}{\right)}
\newcommand{\lk}{\left(}

\newcommand{\lla}{\left\langle}
\newcommand{\rra}{\right\rangle}

\newcommand{\rarr}{\rightarrow}

\newcommand{\bra}[1]{\lla #1 |\right.}
\newcommand{\ket}[1]{\left. | #1 \rra}
\newcommand{\braket}[2]{\ensuremath{\lla{#1} | {#2} \rra}}

\newcommand{\bea}{\begin{eqnarray}}
\newcommand{\eea}{\end{eqnarray}}

\begin{document}

\section{Introduction}

The confinement puzzle has been with us ever since the birth of
quantum chromodynamics (QCD). By means of lattice calculations, it has
been possible to penetrate the infrared nonperturbative sector of QCD
and recover a confining potential between (static) quarks
\cite{Gre03+Bal00}. At present, however,
available lattice sizes do not suffice to describe the Green
functions in the deep infrared \cite{latttorus}.

The continuum approach, on the other hand, has the intriguing feature that
the infrared limit can be studied asymptotically. In the last decade,
a new understanding of infrared QCD has arisen from studying
continuum Yang-Mills (YM) theory via Dyson-Schwinger equations. The Landau gauge has the
advantage of being covariant and therefore encouraged many to
intensive investigation of the infrared properties of YM
theory \cite{SmeAlkHau,Fis06}. In Coulomb gauge, non-covariance brings about severe
technical difficulties which are only recently on the verge of being
overcome \cite{WatRei07}. Nevertheless, the Coulomb gauge might be the more
efficient choice to identify the nonabelian degrees of freedom. It is well-known that screening and
anti-screening contributions to the interquark potential are neatly
separated in Coulomb gauge perturbation theory \cite{Khr70}. As for the infrared
domain, the Gribov-Zwanziger scenario serves as a transparent
confinement mechanism \cite{Gri78,Zwa97}.

A further advantage of working in the physical Coulomb gauge is that
one may pass over to a Hamiltonian description. This opens up direct
access to the heavy quark potential via the expectation value
of the Hamiltonian. In recent years, variational methods have been
pursued to solve the Yang-Mills Schr\"odinger equation with a Gaussian
type of wave functional
\cite{SzcSwa02,FeuRei04,SchLedRei06,EppReiSch07}. Despite Feynman's
critique \cite{ConfWan87}, it turns out that the wave functional is
sensitive to infrared modes and the variational method a powerful
tool, at least for the qualitative description of YM theory. With
careful treatment of the operator ordering in Coulomb gauge
\cite{ChrLee80}, it is possible to find a strictly linearly rising
heavy quark potential. We report on the latest results found in the
Hamiltonian approach to YM theory in Coulomb gauge. This includes the
full calculation of gluon and ghost Green functions and a running
coupling. Furthermore, the 't Hooft loop, an (dis-)order parameter for
confinement, will be calculated using the results of the Green functions.

This paper is organized as follows. In section \ref{setup}, the
Yang-Mills Hamiltonian in Coulomb gauge and the equations of motion are
introduced. The latter will be solved variationally and the heavy quark potential
and the running coupling are presented in section \ref{wdx}. The 't
Hooft loop is discussed in section \ref{hooft} and conclusions are given
in section \ref{summ}.

\section{Yang-Mills Schr\"odinger equation and Dyson-Schwinger equations}
\label{setup}

 In the canonical quantization approach, we choose $A_0^a(x)=0$ and
 impose the usual  equal-time commutation relations among the gauge
 field $A_i^a(x)$ and the conjugate momentum $\Pi_i^a(x)$ to arrive
 at the Weyl gauge Hamiltonian. Since $A_0$ originally serves as the Lagrange
 parameter of the Gauss law, the choice of Weyl gauge requires a
 restriction on the Hilbert space,
 \begin{equation}
   \label{GaussLaw}
   \hat{D}\Pi\ket{\Psi}=g\rho_m\ket{\Psi}
 \end{equation}
where $g$ is the gauge coupling, $\rho_m^a(x)$ the density  of
external color charges, and
$\hat{D}_i^{ab}=\partial_i\delta^{ab}+g\hat{A}_i^{ab}$ with
$\hat{A}_i^{ab}=A^c_if^{acb}$. Fixing the residual time-independent
local gauge invariance by the
Coulomb gauge, $\partial_iA_i=0$, and eliminating the
longitudinal part of the momentum operator $\Pi$ by means of the Gauss law  
(\ref{GaussLaw}), one arrives at the Hamiltonian that depends
only on transversal fields,
\begin{equation}
  \label{Hamiltonian}
  H=\frac{1}{2}\int\lk \cJ^{-1}\Pi\cJ \Pi +B^2+g^2 \cJ^{-1} \rho\: F
  \cJ \rho \rk \; .
\end{equation}
The appearance of the Faddeev-Popov determinant $\cJ[A]=\det(-D[A]\partial)$
is due to a non-trivial change of coordinates to the transverse fields
and turns out to be crucial to the infrared properties of the
theory. The latter term in Eq.\ (\ref{Hamiltonian}) describes the
Coulomb interaction of dynamical and external charges $\rho=-\hat
A\Pi+\rho_m$ via
\begin{equation}
  \label{CoulombOp}
  F^{ab}(x,y)=\bra{x,a}\left( -D\partial \right)^{-1}(-\partial^2)\left( -D\partial \right)^{-1}\ket{y,b}
\end{equation}
and reduces to the familiar Coulomb law in the abelian theory. 

With the Hamiltonian (\ref{Hamiltonian}) at hand, we may apply the
variational principle to find the wave functional
$\Psi[A]=\braket{A}{\Psi}$. Inspired by QED, we choose \cite{FeuRei04}
\begin{equation}
  \label{psi}
  \Psi[A]=\frac{\cN}{\sqrt{\cJ}}\int \cD A \exp\left(-\int A\omega A\right)
\end{equation}
with a normalization
constant $\cN$. The factor of
$\cJ^{-1/2}$ is chosen to alleviate the computation of expectation
values, similar to defining radial states in quantum mechanics. A
different power of $\cJ$ in the wave functional does not change the
properties of the solution \cite{ReiFeu04}. One may think of  the
variational parameter $\omega$ as in the inverse of the gluon
propagator,
\begin{equation}
  \label{gluon}
  D_{ij}^{ab}(x,y)=\bra{\Psi}A_i^a(x)A_j^b(y)\ket{\Psi}=\frac{1}{2}\delta^{ab}t_{ij}(x)\omega^{-1}(x,y)
\end{equation}
with $t_{ij}$ being the transverse projector. It is determined by
solving the functional Schr\"odinger equation, i.e.\ minimizing the
energy $\bra{\Psi}H\ket{\Psi}$. This gives rise to a non-linear integral equation in $\omega$ which we
refer to as the gap equation. It was derived to two-loop order in the
energy in Ref.\ \cite{FeuRei04} and reads in momentum space ($k=|\mathbf{k}|$)
\begin{equation}
  \label{gapeqn}
  \omega^2(k)=k^2+\chi^2(k)+I_\omega(k)+I_\omega^0\; .
\end{equation}
Here, $\chi(k)$ abbreviates the so-called curvature and it is related by
\begin{equation}
  \label{curvature}
  \chi(k)= \frac{N_c}{4}\int \frac{d^3q}{(2\pi)^3} \left(1-(\hat{\mathbf{k}}\cdot\hat{\mathbf{q}})^2\right) \frac{d(|\mathbf{k}-\mathbf{q}|)\:d(q)}{(\mathbf{k}-\mathbf{q})^2}
\end{equation}
to the ghost propagator
\begin{equation}
  \label{ghost}
  \bra{\Psi}\left( -D\partial
\right)^{-1}\ket{\Psi}=\frac{1}{g}\frac{d(k)}{k^2} \; .
\end{equation}
A Dyson-Schwinger equation for the ghost form factor $d$ may be
derived from the path integral, or alternatively from the following operator
identity for $G[A]=\left( -D\partial \right)^{-1}$,
\begin{equation}
  \label{Dyson}
  G[A]=\left( -\partial^2
\right)^{-1}+\left( -\partial^2 \right)^{-1}g\hat A\partial \:G[A]
\end{equation}
which yields
\begin{equation}
  \label{ghosteqn}
  d^{-1}(k)=g^{-1}-\frac{N_c}{2}\int\frac{d^3q}{(2\pi)^3}
  \left(1-(\hat{\mathbf{k}}\cdot\hat{\mathbf{q}})^2\right)\frac{d(|\mathbf{k}-\mathbf{q}|)}{(\mathbf{k}-\mathbf{q})^2\:\omega(q)}\; .
\end{equation}
Both in the ghost Dyson-Schwinger equation (\ref{ghosteqn}) and in the
equation for the curvature (\ref{curvature}), we have approximated the proper ghost-gluon vertex by
its tree-level counterpart $ \Gamma^0_i$. This amounts to the factorization
\begin{equation}
  \label{factor}
  \lla A_i G[A]\rra
= D_{ij} \lla G[A]\: \Gamma^0_j\: G[A]\rra
\approx D_{ij}\lla G[A] \rra \Gamma^0_j\lla G[A]\rra\; .
\end{equation}
The non-renormalization of the ghost-gluon vertex in gauges where the
gluon propagator is transverse, such as the Coulomb and the Landau
gauge \cite{Tay71,FisZwa05}, suggests that the above approximation is
a good one. Dyson-Schwinger studies in both four and three-dimensional
Landau gauge as well as lattice calculations in four-dimensional
Landau gauge confirmed that the dressed vertex is close to
tree-level \cite{Sch+05+CucMenMih04}. The case of
three-dimensional Landau gauge resembles the Coulomb gauge and
therefore we adopt the approximation (\ref{factor}). This vertex'
non-renormalization will have crucial impact on the IR sector of the solutions.

The other momentum dependent term $I_\omega(k)$ in the gap equation (\ref{gapeqn}) reads
\begin{equation}
  \label{Coulombterm}
  I_\omega(k)=\frac{N_C}{4} \int \frac{d^3 q}{(2 \pi)^3} \lk 1 + (\hat{{\bf k}}\cdot \hat{{\bf q}})^2
\rk 
\frac{d ({\bf k} -{\bf q})^2 f ({\bf k} - {\bf q})}{({\bf k} - {\bf q})^2} 
\frac{\left[ \omega ({\bf q}) - \chi ({\bf q}) + \chi
({\bf k}) \right]^2 - \omega ({\bf k})^2}{\omega ({\bf q})} \hk
\end{equation}
and is due to the Coulomb interaction part of the Hamiltonian. Here,
the form factor $f$ measures the deviation from the factorization of
the Coulomb potential, 
\begin{equation}
  \label{fdef}
  \lla G[A](-\partial^2) G[A]\rra = \lla G[A]\rra(-\partial^2)f\lla
  G[A]\rra\; .
\end{equation}
In the infrared, we set $f(k)=1$, factorizing the expectation value for
the Coulomb propagator (\ref{fdef}) equivalently to the one for the
ghost-gluon vertex in Eq.\ (\ref{factor}). In the ultraviolet, $f(k)$ is
treated perturbatively, see \cite{FeuRei04}.

In order to fix the Coulomb gauge uniquely, configuration space must
be restricted to the compact fundamental modular region. As suggested
in \cite{Zwa97}, this entails the ``horizon condition'' for the ghost form
factor,
\begin{equation}
  \label{horizon}
  d^{-1}(0)=0\; .
\end{equation}
As we shall see, the horizon condition (\ref{horizon}) has the consequence that all form factors
$d$, $\chi$ and $\omega$ diverge in the infrared.

\section{Green functions, heavy quark potential and running coupling}
\label{wdx}

The ultraviolet divergences encountered in the gap equation
(\ref{gapeqn}) are removed by subtracting the equations at an arbitrary
renormalization scale $\mu$.  Alternatively, one can eliminate the
divergences by adding appropriate counter terms to the YM Hamiltonian
and to $ln\cJ$ \cite{AdamCorr}. This eliminates the UV-divergent constant $I_\omega^0$
from Eq.\ (\ref{gapeqn}) and involves some renormalization
constants, one of them can be chosen as $c=\lim_{k\rarr
  0}(\omega(k)-\chi(k))$ and fixed by the requirement of minimal energy to be $c=0$. For details,
see Ref.\ \cite{Reinhardt:2007wh}.

\begin{figure}
\begin{center}
\includegraphics[scale=0.96]{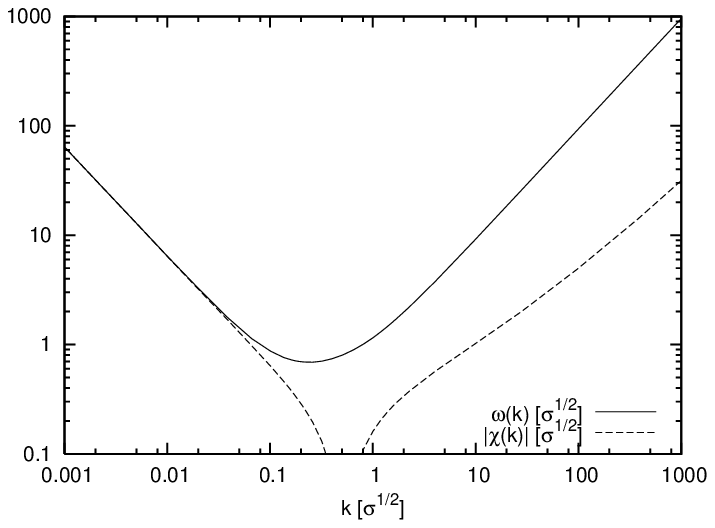}\includegraphics[scale=0.96]{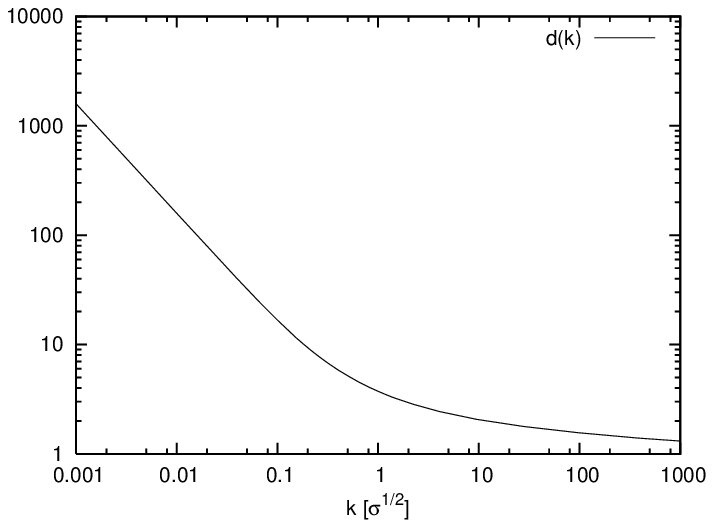}
\end{center}
\caption{Left: the gluon form factor $\omega(k)$ and the
  curvature $|\chi(k)|$. Right: the ghost form factor $d(k)$.
\label{form}}
\end{figure}
The solutions for the form factors $\omega(k)$, $d(k)$ and
$\chi(k)$ can be seen in Fig.\ \ref{form}. In the asymptotic infrared, the
gluon form factor $\omega(k)$ approaches the curvature $\chi(k)$,
reflecting the dominance of ghost degrees of freedom, cf.\ Landau
gauge \cite{Fis06}. With $\chi(k)$ being infrared enhanced, the
ghost content of the solution makes propagation of gluons over
asymptotically large distances impossible, hence gluons are confined.

The correlation of the asymptotic infrared power laws is due to
the non-renormalization of the ghost-gluon vertex,
\begin{equation}
  \label{IRlaws}
  \omega(k)=\chi(k)\sim d(k)\sim\frac{1}{k}\: ,\; f(k)=1\: ,\quad
  k\rarr 0\: .
\end{equation}
Without imposing the horizon condition (\ref{horizon}), solutions to
$d$, $\chi$ and $\omega$ can be found that approach finite values in
the infrared \cite{AdamCorr} so that the energy functional is
dominated by the ultraviolet modes, as speculated by Feynman
\cite{ConfWan87}. Conversely, the infrared power law
solution (\ref{IRlaws}) where the horizon condition is satisfied are
not subdominant to ultraviolet modes and turn out independently of the
details of the wave functional. Even
a stochastic vacuum, $\Psi[A]=1$, would produce the same results for
the infrared \cite{ReiFeu04}. One may therefore be confident using the
variational principle.

The infrared enhancement of the form factors $\omega(k)$ and $d(k)$ is
qualitatively reproduced by recent lattice calculations \cite{PoS}. 

\begin{figure}
\begin{center}
\includegraphics[scale=0.96]{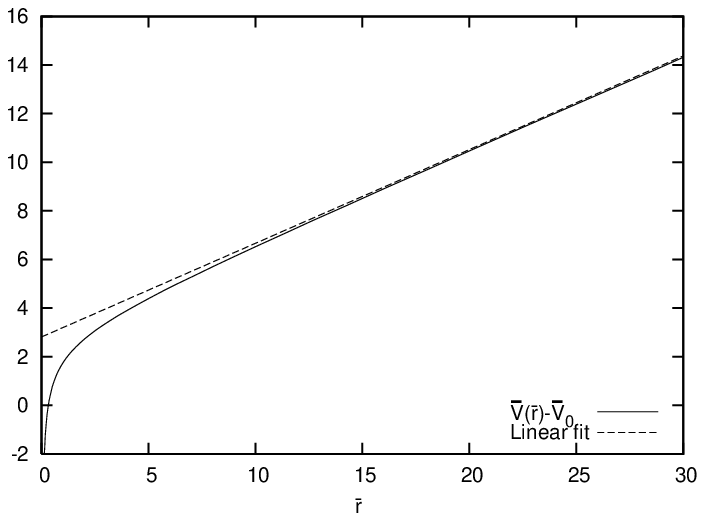}\includegraphics[scale=0.96]{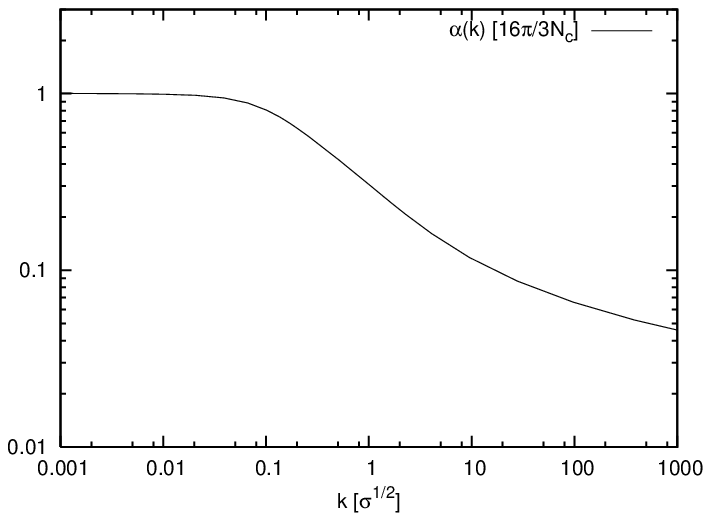}
\end{center}
\caption{Right: the Coulomb
potential $V(r)$. Right: the running coupling $\alpha(k)$.
\label{fig-alpha-quark}}
\end{figure}

Equipped with the ghost form factor $d(k)$, the heavy quark potential
can be found by choosing 
\begin{equation}
  \label{rhom}
  \rho_m^a(x)=\delta^{a3}\left(\delta^{(3)}(x-r/2)-\delta^{(3)}(x+r/2).\right)
\end{equation}
and recalculating the energy $\lla H \rra$ with fixed $\omega$. There is only one
contribution to the energy that depends on the distance $r$ between
the quarks. Using Eq.\ (\ref{fdef}), it reads
\begin{equation}
  \label{potential}
  V_c(r)=\frac{g^2}{2}\int\bra{\Psi}\rho_m \: F \: \rho_m\ket{\Psi}=\int \frac{d^3q}{(2\pi)^3}
\frac{d^2(q)f(q)}{q^2}\left(1-e^{i\mathbf{q}\cdot\mathbf{r}}\right)\, .
\end{equation}
With the infrared behavior of the form factors (\ref{IRlaws}), we
find that $V_c(r)$ rises linearly in the infrared and thus confines
heavy quarks as shown in Fig.\ \ref{fig-alpha-quark}. By matching the
slope of the linear potential to the lattice string tension $\sigma$,
one may set the scale.

Apart from the solution in Fig.\ \ref{form} with the asymptotic
infrared behavior (\ref{IRlaws}) there is one further solution with slightly different infrared exponents for the power laws. The
latter was discovered first \cite{FeuRei04}, however, it does
not have the same attractive features as the one in Fig.\
\ref{form}. In particular, the heavy quark potential is strictly
linearly rising only for the solution presented here. 

A nonperturbative running coupling may be extracted from the
ghost-gluon vertex \cite{FisZwa05,SchLedRei06},
\begin{equation}
  \label{runncoup}
  \alpha(k)=\frac{2}{3\pi}k\,d^2(k)\,\omega^{-1}(k)\, ,
\end{equation}
With a tree-level vertex, it can be shown that one finds a finite value
in the infrared, $\alpha(0)=16\pi/(3N_c)$ \cite{SchLedRei06}. In the ultraviolet, we find
the correct $1/\ln(k/\mu)$ scaling from one-loop perturbation
theory. However, the first coefficient of the beta function,
$\beta_0$, is off by a factor of $8/11$. This is due to the
approximations and requires further investigation. Recent lattice
calculations in Coulomb gauge \cite{FurNak07} show qualitative agreement with the running coupling
presented here. In comparison to
the analogous running coupling in the Landau gauge \cite{Fis06}, we do not find a
bump for intermediate momenta yielding a spurious zero in the beta
function. Note that $\alpha(k)$ in Fig.\ \ref{fig-alpha-quark} is a monotonic function. 
\section{The 't Hooft loop}
\label{hooft}
A (dis-)order parameter of confinement is the 't Hooft loop $\langle V (C)
\rangle$ \cite{'t Hooft:1977hy} whose operator $V (C)$ is defined by the 
relation $V (C_1) W
(C_2) = Z^{L (C_1, C_2)} W (C_2) V (C_1)$, where $W (C)$ is the operator of the
spatial Wilson loop, $Z$ is a (non-trivial) center element of the gauge group
and $L (C_1, C_2)$ denotes the Gaussian linking number. An explicit realization
of $V (C)$ in continuum Yang-Mills theory was derived in ref.
\cite{Reinhardt:2002mb}  and is
given by
\be
\label{G1}
V (C) = \exp \left[ i \int d^3 x \cA^a_i [C] (x) \Pi^a_i (x) \right] \hk .
\ee
Here $\cA [C]$ denotes the gauge potential of a (spatial) center vortex whose
magnetic flux is localized at the loop $C$. Since $V (C) \Psi (A) = \Psi (A +
\cA [C])$ the 't Hooft loop is a center vortex generator. Using the wave
functional found in the variational solution of the Yang-Mills Schr\"odinger
equation in Coulomb gauge, as described above, the expectation value $\langle V
(C) \rangle \equiv \exp (- S (C))$  was evaluated for a planar circular loop
$C$ and is was found that the exponential $S (C)$ obeys a perimeter law
signaling confinement \cite{Reinhardt:2007wh}. This result is in accord with the linear behavior found
for the static color potential.
\section{Conclusions}
\label{summ}
We have solved the Yang-Mills Schr\"odinger equation approximately and
thus determined the vacuum wave functional. Our solutions exhibit the
phenomena of confinement of gluons as well as heavy quarks. As an
improvement on previous results, the heavy quark potential rises
strictly linearly. The nonperturbative running coupling derived from
the ghost-gluon vertex was presented and the 't Hooft loop was
calculated. It is promising that the results have the crucial features
of nonperturbative physics, and that calls for further investigations in
the Hamiltonian approach.
\section{Acknowledgments}
It is a pleasure to thank A.\ Szczepaniak and A.\ Weber for continuing
correspondence on the subject. One of us (H.R.) would like to thank
the organizers of the conference for the stimulating event.


\begin{thebibliography}{99}
\addtolength{\itemsep}{-6pt}
\bibitem{Gre03+Bal00}
  J.~Greensite,
  Prog.\ Part.\ Nucl.\ Phys.\  {\bf 51} (2003) 1
  [arXiv:hep-lat/0301023];
  G.~S.~Bali,
  Phys.\ Rept.\  {\bf 343} (2001) 1
  [arXiv:hep-ph/0001312].
\bibitem{latttorus}
  C.~S.~Fischer, R.~Alkofer and H.~Reinhardt,
  Phys.\ Rev.\  D {\bf 65} (2002) 094008
  [arXiv:hep-ph/0202195];
  C.~S.~Fischer, A.~Maas, J.~M.~Pawlowski and L.~von Smekal,
  arXiv:hep-ph/0701050.
\bibitem{SmeAlkHau}
  L.~von Smekal, R.~Alkofer and A.~Hauck,
  Phys.\ Rev.\ Lett.\  {\bf 79} (1997) 3591
  [arXiv:hep-ph/9705242];
  L.~von Smekal, A.~Hauck and R.~Alkofer,
  Annals Phys.\  {\bf 267} (1998) 1
  [Erratum-ibid.\  {\bf 269} (1998) 182]
  [arXiv:hep-ph/9707327].
\bibitem{Fis06}
  C.~S.~Fischer,
  J.\ Phys.\ G {\bf 32} (2006) R253
  [arXiv:hep-ph/0605173], and references therein.
\bibitem{WatRei07}
    P.~Watson and H.~Reinhardt,
  arXiv:0709.3963 [hep-th],
  arXiv:0709.0140 [hep-th];
  Phys.\ Rev.\  D {\bf 75} (2007) 045021
  [arXiv:hep-th/0612114].
\bibitem{Khr70}
  I.~B.~Khriplovich,
  Yad.\ Fiz.\  {\bf 10} (1969) 409.
\bibitem{Gri78}
  V.~N.~Gribov,
  Nucl.\ Phys.\  B {\bf 139} (1978) 1.
\bibitem{Zwa97}
  D.~Zwanziger,
  Nucl.\ Phys.\  B {\bf 485} (1997) 185
  [arXiv:hep-th/9603203].
\bibitem{SzcSwa02}
  A.~P.~Szczepaniak and E.~S.~Swanson,
  Phys.\ Rev.\ D {\bf 65} (2001) 025012
  [arXiv:hep-ph/0107078];
  A.~P.~Szczepaniak,
  Phys.\ Rev.\ D {\bf 69} (2004) 074031
  [arXiv:hep-ph/0306030].
\bibitem{FeuRei04}
  C.~Feuchter and H.~Reinhardt,
  Phys.\ Rev.\ D {\bf 70} (2004) 105021
  [arXiv:hep-th/0408236], 
  arXiv:hep-th/0402106.
\bibitem{SchLedRei06}
  W.~Schleifenbaum, M.~Leder and H.~Reinhardt,
  Phys.\ Rev.\ D {\bf 73} (2006) 125019
  [arXiv:hep-th/0605115].
\bibitem{EppReiSch07}
  D.~Epple, H.~Reinhardt and W.~Schleifenbaum,
  Phys.\ Rev.\  D {\bf 75} (2007) 045011
  [arXiv:hep-th/0612241].
\bibitem{ConfWan87}
  R.~P.~Feynman,
in {\it *Wangerooge 1987, proceedings, Variational Calculations in
  Quantum Field Theory*}, p.\ 28-40. 
\bibitem{ChrLee80}
  N.~H.~Christ and T.~D.~Lee,
  Phys.\ Rev.\ D {\bf 22} (1980) 939
  [Phys.\ Scripta {\bf 23} (1981) 970].
\bibitem{ReiFeu04}
  H.~Reinhardt and C.~Feuchter,
  Phys.\ Rev.\ D {\bf 71} (2005) 105002
  [arXiv:hep-th/0408237].
\bibitem{Tay71}
  J.~C.~Taylor,
  Nucl.\ Phys.\ B {\bf 33} (1971) 436; 
\bibitem{FisZwa05}
  C.~S.~Fischer and D.~Zwanziger,
  Phys.\ Rev.\ D {\bf 72} (2005) 054005
  [arXiv:hep-ph/0504244].
\bibitem{Sch+05+CucMenMih04}
  A.~Cucchieri, T.~Mendes and A.~Mihara,
  JHEP {\bf 0412} (2004) 012
  [arXiv:hep-lat/0408034];
  W.~Schleifenbaum, A.~Maas, J.~Wambach and R.~Alkofer, Phys.\ Rev.\ D
  {\bf 72} (2005) 014017 [arXiv:hep-ph/0411052];
  A.~Sternbeck, E.~M.~Ilgenfritz, M.~Muller-Preussker and A.~Schiller,
  PoS {\bf LAT2005} (2006) 333
  [arXiv:hep-lat/0509090].
\bibitem{AdamCorr}
  D.\ Epple, H.\ Reinhardt, W.\ Schleifenbaum,\ A. Szczepaniak, in preparation.
\bibitem{PoS}
M.\ Quandt, PoS {\bf LAT2007} 325;
  A.~Voigt, E.~M.~Ilgenfritz, M.~Mueller-Preussker and A.~Sternbeck,
  arXiv:0709.4585 [hep-lat], PoS {\bf LAT2007} 338.
\bibitem{FurNak07}
  S.~Furui and H.~Nakajima,
  arXiv:0708.1421 [hep-lat], PoS {\bf LAT2007} 301.
\bibitem{'t Hooft:1977hy}
  G.~'t Hooft,
  Nucl.\ Phys.\  B {\bf 138}, 1 (1978).
\bibitem{Reinhardt:2002mb}
  H.~Reinhardt,
  Phys.\ Lett.\  B {\bf 557}, 317 (2003)
  [arXiv:hep-th/0212264].
\bibitem{Reinhardt:2007wh}
  H.~Reinhardt and D.~Epple,
  arXiv:0706.0175 [hep-th], Phys.\ Rev.\ D, in press.
\end{thebibliography}
\end{document}